\documentclass[11pt]{article}
\usepackage{graphicx}
\usepackage{jheppub} 
\usepackage[T1]{fontenc} 
\usepackage{multirow}    
\usepackage{comment}
\usepackage{amsmath}
\usepackage{amsfonts}
\usepackage{amssymb}
\usepackage{breqn}
\usepackage{bbm}
\usepackage{graphicx}
\usepackage{caption}
\usepackage[usenames, dvipsnames]{color}
\usepackage{slashed}
\usepackage{xcolor}
\usepackage{changepage}

\def\beq{\begin{equation}}
\def\eeq{\end{equation}}

\newcommand{\cm}{\rm{cm}}
\newcommand{\bea}{\begin{eqnarray}\begin{aligned}}
\newcommand{\eea}{\end{aligned}\end{eqnarray}}
\newcommand{\bmtx}{\begin{pmatrix}}
\newcommand{\emtx}{\end{pmatrix}}

\newcommand{\chis}{\ensuremath{\chi^*}}

\newcommand{\ev}{\text{eV}}
\newcommand{\tev}{\text{TeV}}
\newcommand{\kev}{\textrm{keV}}
\newcommand{\mev}{\text{MeV}}
\newcommand{\gev}{\text{GeV}}
\newcommand{\x}{\chi}
\newcommand{\xs}{\chi^*}
\newcommand{\mx}{m_{\chi}}
\newcommand{\Ap}{A^\prime}
\newcommand{\mAp}{m_{A^\prime}}
\newcommand{\vev}[1]{{\langle #1 \rangle}}

\title{Electromagnetic Signals of Inelastic Dark Matter Scattering}

\author[1,2]{Masha Baryakhtar,}

\author[1]{Asher Berlin,}

\author[1,3]{Hongwan Liu,}

\author[1]{Neal Weiner}
\affiliation[1]{Center for Cosmology and Particle Physics, Department of Physics, New York University, New York, NY 10003, USA}
\affiliation[2]{Department of Physics, University of Washington, Seattle WA 98195, USA}
\affiliation[3]{Department of Physics, Princeton University, Princeton, NJ 08544, USA}

\emailAdd{mbaryakh@uw.edu}
\emailAdd{ajb643@nyu.edu}
\emailAdd{hongwanl@princeton.edu}
\emailAdd{neal.weiner@nyu.edu}
\date{\today}

\abstract{Light dark sectors in thermal contact with the Standard Model naturally produce the observed relic dark matter abundance and are the targets of a broad experimental search program.   A key light dark sector model is the pseudo-Dirac fermion with a dark photon mediator. The dynamics of the fermionic excited states are often neglected. We consider scenarios in which a nontrivial abundance of excited states is produced and their subsequent de-excitation yields interesting electromagnetic signals in direct detection experiments. We study three mechanisms of populating the excited state: a primordial excited fraction, a component up-scattered in the sun, and a component up-scattered in the Earth. We find that the fractional abundance of primordial excited states is generically depleted to exponentially small fractions in the early universe. Nonetheless, this abundance can produce observable signals in current dark matter searches. MeV-scale dark matter with thermal cross sections and higher can be probed by down-scattering following excitation in the sun. Up-scatters of GeV-scale dark matter in the Earth can give rise to signals in current and upcoming terrestrial experiments and X-ray observations. We comment on the possible relevance of these scenarios to the recent excess in XENON1T.}

\begin{document}

\maketitle
\flushbottom

\section{Introduction}

Over the past two decades, the model space of WIMP-like dark matter (DM) has broadened to include new states and interactions. In particular, the introduction of new, light mediators---such as the dark photon---between the dark sector and the Standard Model (SM) has opened the parameter space of  light thermal DM \cite{Alexander:2016aln}.

Light Dirac fermion DM is in severe conflict with measurements of the CMB if it can annihilate at the time of recombination \cite{Padmanabhan:2005es,Slatyer:2009yq,Madhavacheril:2013cna}. However, a pseudo-Dirac fermion avoids the CMB constraints if only the ground state has a significant abundance. Thus, a broad class of viable light DM models necessarily is described by DM with a ground and an excited state, and a light dark photon mediator to the SM \cite{Alexander:2016aln}.

The existence of an excited state $\xs$ of the  DM particle $\x$ can dramatically alter the energy spectrum in direct detection experiments through endothermic \cite{TuckerSmith:2001hy} or exothermic scattering of long-lived states \cite{Finkbeiner:2009mi,Batell:2009vb,Lang:2010cd,Graham:2010ca}.  Decays of the excited state  have important implications for direct  \cite{Finkbeiner:2009ug,Chang:2010en,Feldstein:2010su}, indirect, and accelerator signals~\cite{Morrissey:2014yma,Izaguirre:2015zva,Izaguirre:2017bqb,Berlin:2018jbm,Izaguirre:2015yja}. If the splitting is present, elastic scattering is highly suppressed in light DM models with dark photon mediators. As a consequence, understanding  the possible presence of these excited states and their signals is imperative.

For light DM with mass $m_\chi $ at the MeV scale, the mass splitting $\delta= m_\chis-m_\chi$ in the dark sector has two natural values. The first is an $\mathcal{O}(1)$ splitting, $\delta\sim\mathcal{O}(\text{MeV})$, which decouples the excited states from questions of direct detection. However, the splitting breaks a symmetry of the theory and can be parametrically smaller than the other scales in the theory.  Thus a second possibility is splittings in the $\delta\sim\alpha/4\pi \times \mathcal{O}(\text{MeV}) \sim \mathcal{O}(\text{keV})$ range, which lead to signatures in direct detection experiments. While nuclear signals of these excited states have been studied \cite{Graham:2010ca,Fox:2013pia,Frandsen:2014ima}, the electronic signals have only recently started to be explored \cite{Bernal:2017mqb,Bloch:2020uzh}. A systematic study of this parameter space and cosmological history, including signals of nuclear and electronic recoils from primordial states, has recently been studied in Ref.~\cite{Gonzalez:2021kzi}. 

In this paper, we will show that ongoing direct detection experiments are sensitive to light thermal relics through their inelastic scattering, filling an important gap in light DM parameter space. In the elastic limit, recoils of MeV-scale particles typically deposit $\sim m_\chi v^2  \sim 10 \,\ev$ of energy \cite{Essig:2011nj,Essig:2015cda,Essig:2017kqs}. Such small energies are difficult to detect, and require  precision experiments with small target masses \cite{Barak:2020fql,Aguilar-Arevalo:2019wdi} or large targets at the expense of reduced background rejection \cite{Akerib:2017uem,Ren:2018gyx,Aprile:2019xxb}. On the other hand, excited states with large, order-MeV splittings typically decay promptly into $e^+e^-$ pairs and have no relevant local abundance.  

Intermediate, keV-scale excitations are potentially long-lived and allow detection in higher-threshold, tonne-scale experiments, including the world-leading xenon experiments \cite{Fu:2017lfc,Akerib:2017uem,Aprile:2020tmw}. It is this latter case that we consider in this paper, with an emphasis on electronic signals of the excited state $\xs$ down-scattering in direct detection experiments. The scenarios we consider provide possible explanations to the excess of electron recoil events reported by the XENON1T collaboration \cite{Aprile:2020tmw}  and predict future testable experimental consequences.

The layout of this paper is as follows: in Sec.~\ref{sec:modelspace}, we present the parameter space of the model in question. In Secs.~\ref{sec:universe}-\ref{sec:earth}, we consider three possible sources of excited states: the early universe, the Sun, and the Earth, and study the  implications of these states in direct detection experiments. We discuss current bounds and future signatures in laboratory and astrophysical observations; and we conclude in Sec.~\ref{sec:discussion}.

\section{Model Space}
\label{sec:modelspace}

We focus on a specific framework of thermal relics, which can make up all of the DM, or be a subcomponent. We consider light pseudo-Dirac fermion  DM particle, with ground state $\chi$ and excited state $\chis$ split by an amount $\delta$, and overall mass scale $m_\x$. The DM possesses a dipole moment and couples to the Standard Model section through a kinetically mixed massive dark photon of mass $m_{A'}$.  This model combines features discussed broadly in the literature \cite{Holdom:1985ag,Boehm:2003hm,Finkbeiner:2007kk,ArkaniHamed:2008qn,Pospelov:2008jd,Hooper:2008im,Knapen:2017xzo,Cohen:2010kn}.  The interaction Lagrangian is,

\bea
{\cal L} \supset \frac{\epsilon}{2} \, F_{\mu\nu} F^{\prime \mu\nu} + i e_D \,\Ap_\mu \, \bar{\x}^* \gamma^\mu \x +  \frac{1}{\Lambda_d} ~ \bar{\chi}^* \sigma^{\mu \nu} \chi \,  F_{\mu\nu}
\, ,
\label{eq:lagrangian}
\eea
where $e_D = \sqrt{4 \pi \alpha_D}$ is the dark photon gauge coupling. Both vector interactions as well as dipole operators are off-diagonal between the mass eigenstates of the pseudo-Dirac fermion. 

The dark photon and dipole can each allow for transitions between the ground and excited DM state. 
In the parameter range  we consider, the decay time through the dark photon is longer than the age of the universe \cite{Finkbeiner:2009mi,Batell:2009vb}. Only once the splitting is $\mathcal{O}(\mev)$ and de-excitations into $e^+e^-$ pairs are allowed, does the lifetime become short. 

The higher-dimensional dipole operator in Eq.~\eqref{eq:lagrangian} can also result in up- and down- scattering, and scenarios involving this have been discussed previously \cite{Chang:2010en,Feldstein:2010su}. The excited state decays at a rate of
\beq\tau^{-1} \sim\pi \delta^3 /\Lambda_d^2 \sim \sec^{-1} \times \left(\frac{\delta}{\kev}\right)^3\left(\frac{\tev}{\Lambda_d}\right)^2.
\eeq
The scale of the dipole moment is important in determining the possible source of the excited states. In principle, a  dark photon-interacting thermal relic need have no dipole operator with electromagnetism at all. A Planck-suppressed dipole operator (i.e., $\Lambda_d \sim M_{\text{pl}}$) does not mediate a decay over the age of the universe for $\delta \lesssim \mev$. Thus, a natural starting point would be to consider  excited states which are produced primordially and are stable on cosmological timescales, e.g.~\cite{Finkbeiner:2009mi,Batell:2009vb}. 

In the presence of a larger dipole moment (or another means of decay), the primordial excited abundance is depleted and local mechanisms for inelastic up-scattering become an important source for excited states. The Sun, given its relatively high temperature, is a natural source, as is the Earth for large enough DM kinetic energy. Excited states from the Sun can lead to direct detection signals on Earth if their lifetime is longer than $\text{AU}/ v \gtrsim 10^5\,\text{sec}$, where $v > v_{\textrm{DM}}$ is the typical $\xs$ velocity after solar scattering. For up-scattering in the Earth, the lifetime can be even shorter; scenarios with very short ($\sim 100 \ \text{$\mu$sec}$) lifetimes have been considered \cite{Chang:2010en}. We will focus on lifetimes greater than $r_{\text{earth}}/v_{\textrm{DM}} \sim 100  \, \text{sec}$ where the entire Earth is a source of excited states. These lifetimes are achieved for dipole moment scale $\Lambda_d \gtrsim (10 - 100) \,\tev$---a relatively mild constraint given that radiative dipoles are typically suppressed by a small charge and a high mass scale.

The inelastic DM-electron scattering mediated by the dark photon is parametrized by the cross-section in the elastic limit,
\beq
\sigma_{\text{el}} =\frac{16 \pi \, \alpha  \, \alpha_D \, \mu^2 \, \epsilon ^2}{\mAp^4}=\frac{16 \pi \,  \, \alpha \, \mu^2 \, y}{m_\chi^4}\, ,
\label{eq:DPxsec}
\eeq
where $\mu$ is the reduced mass of the dark matter-target system, and  $y \equiv \epsilon^2 \, \alpha_D \, (m_\x/\mAp)^4$ is a standard combination of  DM parameters~\cite{Izaguirre:2015yja}.

When the splitting is small compared to the typical scattering energy, the elastic cross-section fully describes the up- and down-scattering. When the splitting becomes significant compared to the overall kinematics, the width of the recoil energy distribution is corrected as
\beq
\Delta E_R =\Delta E_{\rm R,elastic} \sqrt{1\pm\frac{2 \delta}{\mu v^2}}= \frac{2 \mu ^2 v^2 \sqrt{1\pm\frac{2 \delta}{{\mu v}^2}}}{{m_T}}
\, ,
\eeq
compared to the elastic case, where $+(-)$ is for exothermic (endothermic) inelastic scattering, $m_T$ is the target mass, and $v$ is the DM velocity. The overall scattering cross section then becomes (see, e.g., Refs.~\cite{TuckerSmith:2001hy,Lang:2010cd,Finkbeiner:2009mi,Graham:2010ca}),
\beq
\sigma_{\text{inel}}=\sigma_{\text{el}} \,  \sqrt{1\pm\frac{2 \delta}{\mu v^2}} \, .
\label{eq:kinenhance}
\eeq

Since we will be focused on electromagnetic direct detection signals, let us take a moment to describe the scattering and our simplifying assumptions. Given that the focus of our analysis is on DM masses above an MeV--- to retail a consistent cosmology---and splittings above a keV ---which can be observed in large-volume detectors---,   we estimate the event rates in a xenon detector by assuming the $n=4, 5$ orbitals are accessible and populated by ``free'' electrons at rest, and ignore the $n=1,2,3$ orbitals, which are more tightly bound. We expect this to be a good approximation for  $\delta \gtrsim \text{keV}, m_\x \gg \text{MeV}$ and anticipate $\mathcal{O}(1)$ corrections, in particular a broadening of the energy recoil spectrum, for DM masses $\sim$~MeV; a detailed analysis of the electron response can be found in \cite{Bloch:2020uzh,Ema:2020fit}.

We can estimate the validity of this free electron approximation as follows.  In a two-to-two scattering  the electron acquires a final momentum ${|k'|} = \sqrt{2m_e E_R}$, and recoil energy 
\beq E_R = \vec{q}\cdot \vec{v}_{\chi} -\frac{q^2}{2m_\x}+\delta\label{ER}\,,\eeq
 where  $q$ is the momentum transfer.  In the elastic limit $\delta \rightarrow 0$,  to achieve recoil energies above $\sim \text{keV}$ thresholds, one can see that momentum transfers $q\gtrsim $~MeV are required: much larger than both the outgoing electron momentum $\sqrt{2 m_e E_R}$ and the {\it{typical} } bound electron momentum $k_e \sim Z_{\text{eff}}\,\alpha \, m_e \sim $ few keV \cite{Essig:2015cda}. Thus in the elastic limit, the approximation of electrons at rest is badly violated: the initial electron momentum comes from the high-momentum, low probability tail of the electron wavefunctions to achieve large momentum transfers and recoil energies above threshold. Incredible progress in detailed calculations, including relativistic corrections to the high-momentum tails of the electron wavefunctions has been achieved, e.g.~\cite{Essig:2011nj,Essig:2012yx,Essig:2015cda,Roberts:2016xfw,Essig:2017kqs,Catena:2019gfa}. 
 
In contrast, in down-scattering with large enough splitting, the recoil energy is largely dominated by the splitting $\delta$, and the initial electron momentum can be small compared to the momentum transfer $q \sim\sqrt{2 m_e E_R}$. The primary support of the process then comes from the peak of the electron form factor, and the rates are approximately unsuppressed relative to free electron scattering. 
  
 Considering our energy recoil distribution in more detail, the minimum velocity required to achieve a recoil energy $E_R$ is
  \beq v_{\min} = \left|\frac{E_R-\delta}{q} + \frac{q}{2m_\x}\right| \label{vmin}\,.\eeq 
In the approximation that the electron is initially free and at rest, the momentum transfer is  $q=k'=\sqrt{2m_e E_R}$; the peak of the recoil energy spectrum $E_R^{\text{peak}}$ occurs for $v_{\min} = 0$,
  \beq E_R^{\rm peak} \approx \frac{\mu}{m_e}\delta\,. \eeq
 The fractional spread in the recoil energy due to the initial kinetic energy of the DM  can be approximated by setting $v_{\min} = v_0$,
 \beq\Delta E_{R}/E_R \sim v_0\sqrt{m_e/ \delta}\sim  3\%\, ({\rm keV}/E_R)^{1/2}\,,\eeq
which is small compared to detector resolution, $\sigma(E)/E \sim 30\%  ({\rm keV}/E)^{1/2} + 0.3\%$ \cite{Aprile:2020tmw}.

Incorporating the electron momentum at leading order, the momentum transfer is corrected by $q\approx \sqrt{2m_e E_R}\pm k_e$. Since $k_e \ll \sqrt{2m_e E_R}$, the presence of an initial electron momentum distribution only corrects the peak recoil energy by a small amount. 
 However, due to the low velocity of DM in the galaxy, the electron momentum can be comparable to the DM  momentum at the light end of the mass range we consider, i.e. ${k_e}\sim{m_\x v_\x}$. Given a finite electron momentum distribution with typical momentum $k_e$, the recoil distribution is broadened by a factor of order $(k_e / m_\x)\sqrt{m_e \delta}$, which can be comparable to the broadening due to the DM velocity dispersion $v_0\sqrt{m_e \delta}$ for $m_\x \lesssim $ few MeV. Nevertheless, while the recoil distribution broadens for small dark matter masses, we estimate that it does not exceed the detector energy resolution for $m_\x \gtrsim $~MeV. Note that as we will see, solar up-scattered excited DM particles arrive at the detector with a higher velocity and a broader energy distribution than the primordial DM, so the correction due to the electron momentum in the solar scenario is even less significant.

We also ignore the electron binding energy. The above expressions should be modified to $\delta \rightarrow \delta - E_b$  where $E_b = 12.4 (5p), 25.7  (5s), 75.6 (4d), 163.5 (4p),$ and $213.8 (4s)$~eV  for the outer shells; neglecting the binding energy for the outer shell electrons is valid as $E_b\ll \delta$ in our entire parameter space. The inner shells will be accessible for larger values of the splitting, and the rates would be modified for $\delta \gg {\rm keV}$.

Having laid out the basic ingredients and properties of the inelastic DM model, we now turn to the DM production and detection mechanisms in more detail. In the following sections, we consider three distinct sources of excited states: primordial abundances, excitations from solar reflection, and up-scatterings in the Earth.

\section{Excited States from the Early Universe}
\label{sec:universe}

\subsection{Production}

The high densities and temperatures of the early universe efficiently generate a cosmologically stable excited state abundance. If DM was part of a thermal bath in the primordial universe, chemical equilibrium drives the relative abundances of $\x$ and $\x^*$ to comparable values, as is the case in standard cosmologies of thermal relics in which DM was once in equilibrium with ordinary matter. Once the temperature becomes much smaller than the mass splitting $\delta$, the relative abundance of  the excited state $\xs$ (compared to $\x$) is exponentially suppressed and freezes out. Thus, estimating the primordial fraction of $\xs$ at late times requires tracking the cosmological evolution across the periods of $\x , \xs \leftrightarrow \text{SM}$ chemical and kinetic decoupling, as well as the period of $\x \leftrightarrow \xs$ decoupling. In the following discussion, we give simple analytic expressions for the rates of these  processes;  we use a full numerical analysis in the figures.

In the dark photon model Eq.~(\ref{eq:lagrangian}), DM can maintain chemical equilibrium with the SM bath through coannihilations to electromagnetically charged SM particles $f$, as mediated by the dark photon, $\x \x^* \leftrightarrow \Ap \leftrightarrow f f$. For temperatures $T \ll \mAp$, this process is dominated by the exchange of an off-shell $\Ap$. The total comoving $\x + \xs$ density is dictated by the temperature at which these coannihilations decouple. If decoupling occurs at a temperature much greater than the mass splitting $\delta$, the coannihilations rate scales as $\sigma v \sim \alpha \, y / m_\x^2$. The conserved $\x + \xs$ comoving density is then consistent with the observed DM energy density provided that $\sigma v \sim 1 / (T_{\text{eq}} \, m_{\text{pl}})$, where $T_{\text{eq}} \sim 0.8 \ \text{eV}$ is the temperature at matter-radiation equality and $m_{\text{pl}}$ is the Planck mass, which is equivalent to
\beq
\label{eq:yfreezeout}
y \sim 10^{-10} \times \left( \frac{m_\x}{100 \ \text{MeV}} \right)^2 \, .
\eeq

After chemically decoupling from the SM, $\x$ and $\x^*$ remain chemically coupled to one another through $\xs \xs \leftrightarrow \x \x$ and $\xs f \leftrightarrow \x f$, where the latter process also enforces kinetic equilibrium between the dark sector and the SM. Neither process alters the total $\x + \x^*$ number, but each drives the relative number density to the equilibrium value $n_\xs / n_{\x} \sim e^{- \delta / T_\x}$, where $T_\x$ is the temperature of the $\x + \xs$ bath. Once $\x$ and $\x^*$ chemically decouple from each other, the primordial comoving abundance of the excited state $\xs$ is no longer depleted by annihilation or scattering processes. 

The DM temperature $T_\x$ is governed by the temperature of kinetic decoupling $T_{\text{kin}} \ll m_\x$, which is in turn dictated by DM-electron down-scattering  $\xs e \leftrightarrow \x e$ for $m_\x \lesssim \text{GeV}$. For $m_\x \gg \text{MeV}$, $T \lesssim T_{\text{kin}}$ occurs well after $\x, \xs$ become non-relativistic and chemically decouple from the SM, due to the enhanced abundance of electrons compared to DM particles at early times. In the limit that $m_e \ll T \ll \mAp$, the thermally-averaged rate for $\xs e \leftrightarrow \x e$ is
\beq
\Gamma_{\x e} \simeq \frac{360 \, \zeta(5)}{\pi} ~ \frac{\alpha \, \alpha_D \, \epsilon^2 \,  T^5}{\mAp^4} 
~.
\eeq
At much lower temperatures, $T \ll m_e$, $\Gamma_{\x e}$ is exponentially suppressed, due to the dwindling electron abundance. We estimate $T_{\text{kin}}$ as the temperature at which $\Gamma_{\x e}$  drops below the rate of Hubble expansion $H$. For $T \lesssim T_{\text{kin}}$, the DM temperature evolves independently of the SM plasma as $T_\x \sim T^2 / T_{\text{kin}}$. In most of the parameter space that we investigate, kinetic decoupling occurs near or slightly below the electron mass threshold. 

\begin{figure}\begin{center}
\includegraphics[width=0.495\textwidth]{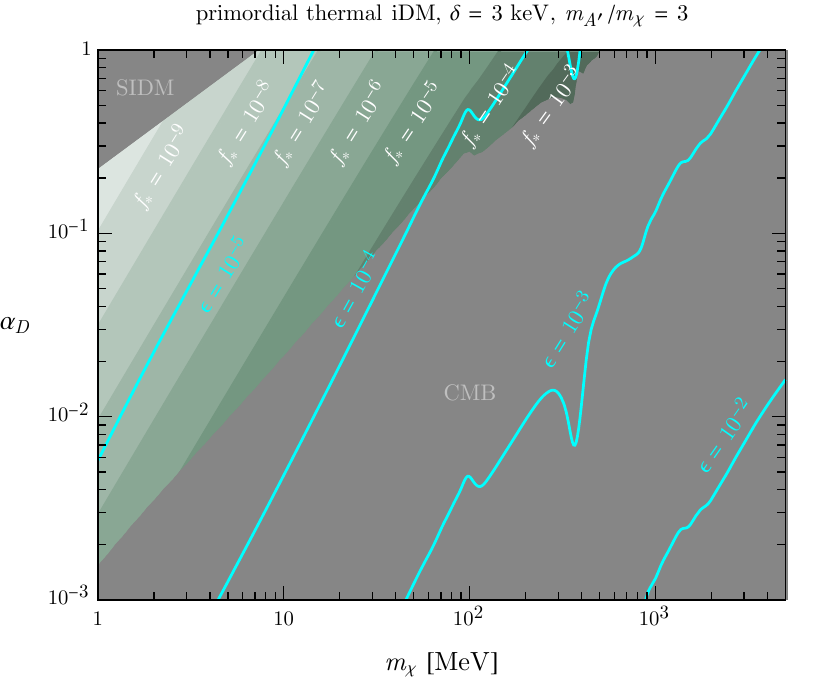} 
\includegraphics[width=0.495\textwidth]{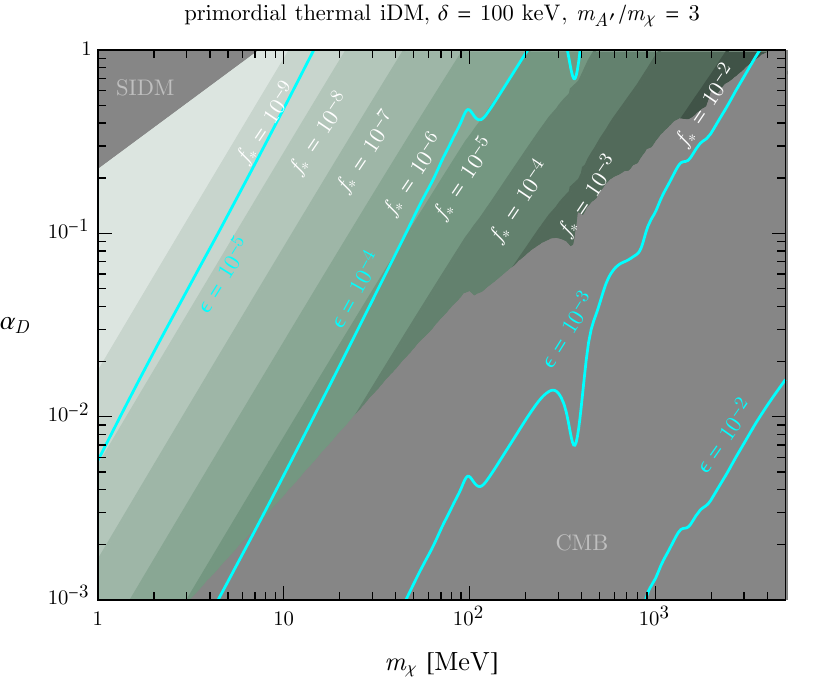}
\caption{The fraction $f_*$ of dark matter that is composed of excited states (shaded green) as a function of dark matter mass $m_\x$ and dark sector coupling $\alpha_D$ for $\mAp = 3 \, m_\x$, with $\xs - \x$ mass splitting $\delta = 3 \ \text{ keV}$ (left) and $100 \text{ keV}$ (right). For each point in parameter space, we fix the kinetic mixing parameter $\epsilon$ such that the abundance of $\x$ agrees with the observed dark matter energy density (cyan). Shown in gray are regions excluded by elastic self-scattering of dark matter~\cite{Berlin:2018jbm,Tulin:2017ara} and distortions of the CMB from late-time annihilations~\cite{Aghanim:2018eyx}.}\label{fig:f2eps}\end{center}
\end{figure}

\begin{figure}\begin{center}
\includegraphics[width=0.495\textwidth]{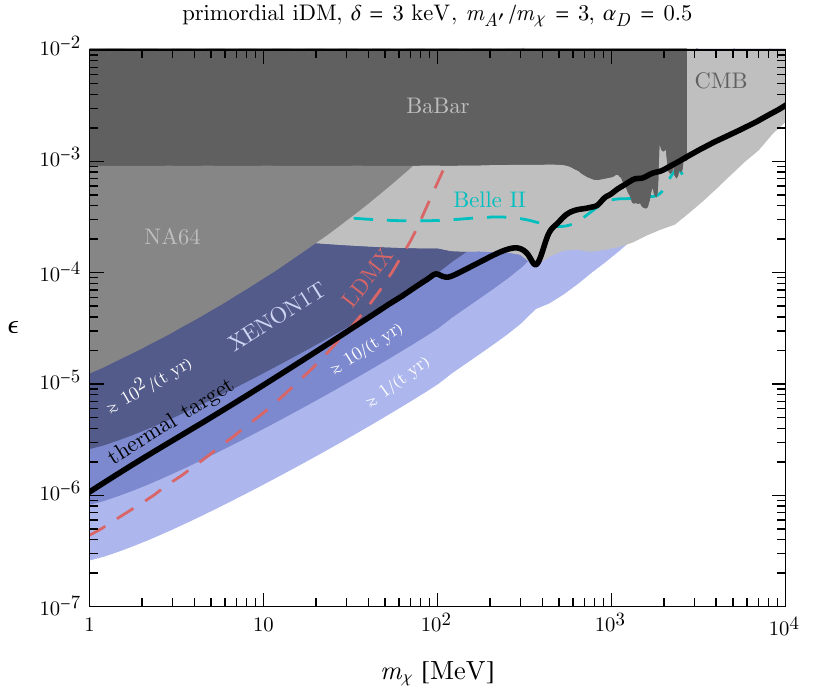} 
\includegraphics[width=0.495\textwidth]{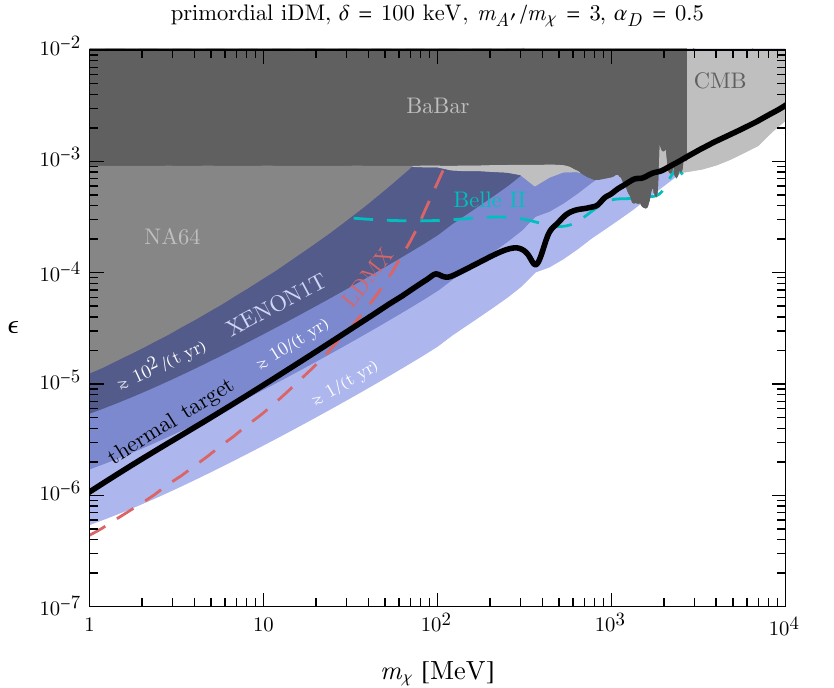}
\caption{In blue, the event yield at XENON1T from down-scattering of a primordial excited pseudo-Dirac dark matter subcomponent as a function of $\epsilon$ and $m_\x$, for $\alpha_D = 0.5$, $\mAp / m_\x = 3$, and mass splitting $\delta = 3 \ \text{ keV}$ (left) and $100 \text{ keV}$ (right).  Throughout, we assume that $\x$ makes up the entirety of the dark matter abundance; along the black contours, the thermal abundance of $\x$ is consistent with the observed dark matter energy density. Also shown are regions excluded by recent missing energy/momentum searches at NA64~\cite{NA64:2019imj} and BaBar~\cite{Lees:2017lec} (solid gray), as well as the projected sensitivities of searches for similar signals at LDMX and Belle II (dashed)~\cite{Izaguirre:2014bca,Battaglieri:2017aum,Akesson:2018vlm,Berlin:2018bsc}. Exclusions derived from distortions of the CMB anisotropies are also shown (solid gray)~\cite{Aghanim:2018eyx}.  Constraints from dark matter self-interactions bounds may apply for $\mx \lesssim 10$~MeV when $\chi$ comprises all of the dark matter.}\label{fig:epsmx}\end{center}
\end{figure}

Even at temperatures well below the electron threshold, $\x$ and $\xs$ can remain in chemical equilibrium through DM-DM scattering $\xs \xs \leftrightarrow \x \x$, which is independent of $\epsilon$. Assuming that $\x$ and $\xs$ are chemically coupled, $n_{\xs} \sim e^{-\delta / T_\x} \, n_\x$. The corresponding thermally-averaged rate is roughly
\beq
\Gamma_{\xs \x} \simeq \, e^{- \delta / T_\x} \, n_\x \, \frac{2^{5/2} \pi \, \alpha_D^2 \, m_\x^{3/2}}{\mAp^4} ~ \text{max} \left(\frac{2}{\pi} \, T_\x \, , \, \delta \right)^{1/2}
~.
\eeq
We denote the DM temperature at which $\Gamma_{\xs \x} \sim H$  as $T_{\x \xs}$, which we evaluate numerically. Since $\xs e \leftrightarrow \x e$ also enforces $\x - \xs$ chemical equilibrium, the DM temperature of $\x - \xs$ chemical decoupling is $T_{\x, \text{chem}} \sim \text{min}(T_{\text{kin}}, T_{\x \xs})$ and is thus controlled by whichever process, $\xs e \leftrightarrow \x e$ or $\xs \xs \leftrightarrow \x \x$, decouples later. Assuming that $\x$ makes up the dominant component of the DM abundance at late times, the number density $n_\x$ in the expression above corresponds to $n_\x \sim T_{\text{eq}} \, T^3 / m_\x$. If $\xs$ is cosmologically stable, its late-time fractional abundance is then approximated by
\beq
f_* \equiv \frac{n_{\xs}}{n_\x + n_{\xs}} \simeq e^{-\delta / T_{ \x, \text{chem}}}
~.
\eeq
Ignoring the $m_\x$-dependence of $T_{\text{kin}} \sim m_e$ and taking $T_{\x, \text{chem}} \lesssim \delta$, $f_*$ then scales as
\begin{align}
f_*  &\sim \frac{m_\x^{7/2}}{\alpha_D^2 \,  (T_{\x, \text{chem}}  \, \delta)^{1/2}} ~ \frac{(\mAp / m_\x)^4}{m_e^{1/2} \, T_{\text{eq}} \, m_{\text{pl}}} 
\nonumber \\
&\sim \text{few} \times 10^{-4} \times \left( \frac{m_\x}{100 \ \text{MeV}} \right)^{7/2} \left( \frac{\alpha_D}{0.5} \right)^{-2} \left( \frac{\delta}{\text{keV}} \right)^{-1} \left( \frac{T_{\x, \text{chem}}}{\delta} \right)^{-1/2} \left( \frac{\mAp / m_\x}{3} \right)^{4}
\end{align}
for $m_\x \sim \mathcal{O}(\text{MeV})$,  where the ratio $T_{\x, \text{chem}} / \delta \lesssim 1$ grows logarithmically with increasing $m_\x$.

For DM masses well below the GeV-scale, the remaining fractional abundance of excited states $\xs$ is typically very small, $f_* \ll 1$. The dependence of $f_*$ on various parameters is shown in Fig.~\ref{fig:f2eps}, in which we vary $\epsilon$ as a function of $\alpha_D$ and $m_\x$ by fixing the late-time abundance of $\x$ to the observed DM energy density. From Eq.~(\ref{eq:yfreezeout}), this  ``thermal target''   corresponds to
\beq
\epsilon \sim 10^{-4} \times \left( \frac{m_\x}{100 \ \text{MeV}} \right)  \left( \frac{\mAp / m_\x}{3} \right)^{2} \left( \frac{\alpha_D}{0.5} \right)^{-1/2}
~,
\eeq
also shown as the black contours in the $\epsilon-m_\x$ parameter space of Fig.~\ref{fig:epsmx}. As discussed above, the total late-time abundance is driven by $\x \xs \leftrightarrow ff$ freeze-out, and in our numerical analysis, we include the effect of hadronic resonances and final states~\cite{Izaguirre:2015zva}. For smaller $\alpha_D$ or larger $m_\x$, the ability to deplete the primordial $\xs$ abundance diminishes, leading to an increased primordial excited state fraction $f_*$. For $m_\x \gtrsim \text{few} \times \text{GeV}$, $\xs$ constitutes an $\mathcal{O}(1)$ fraction of the DM density.

\begin{figure}
\begin{center}
\includegraphics[width=0.495\textwidth]{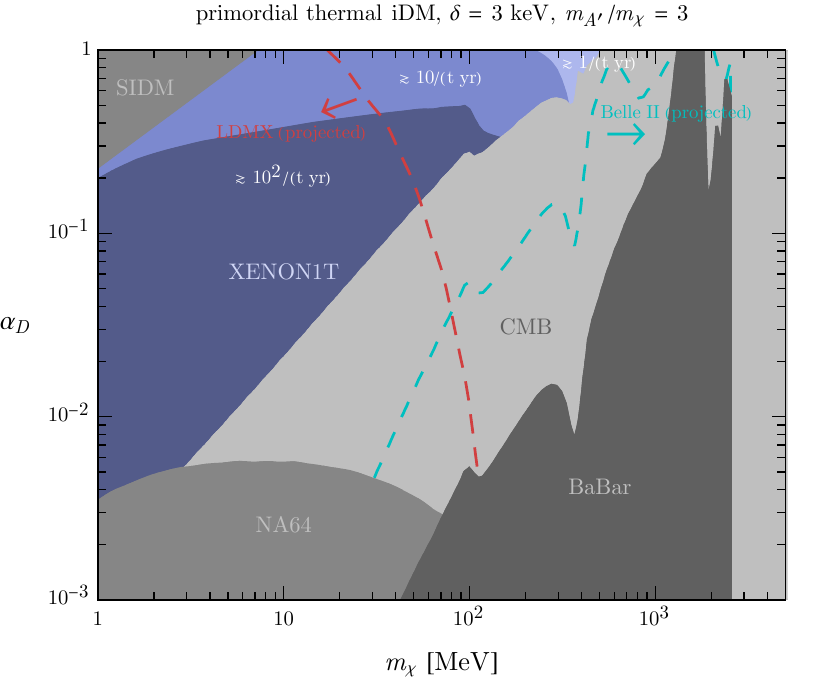} 
\includegraphics[width=0.495\textwidth]{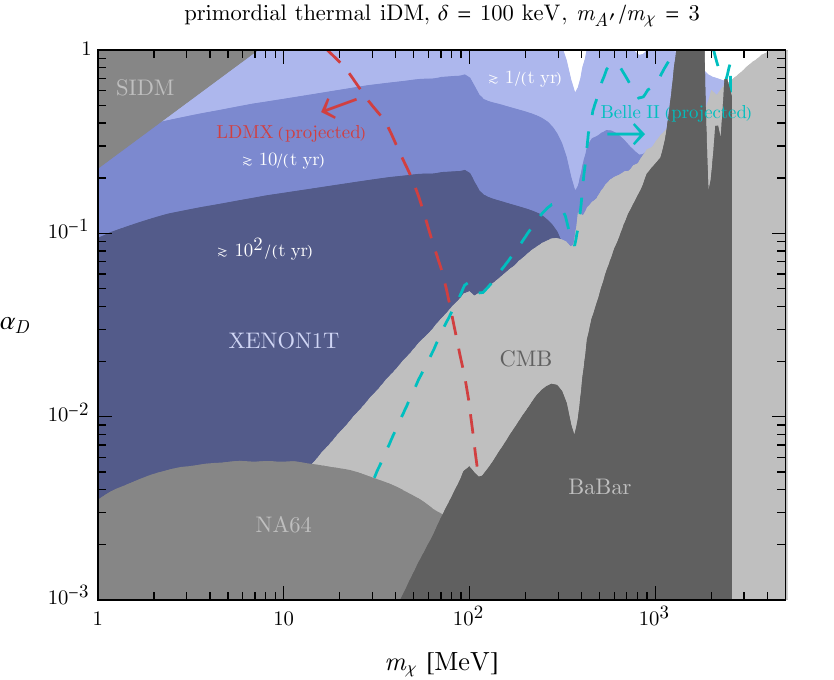}
\caption{As in Fig.~\ref{fig:epsmx}, but now in the $\alpha_D - m_\x$ plane. At each point in parameter space, the value of $\epsilon$ is fixed such that $\x$ freezes out with an abundance that is consistent with the observed dark matter energy density, as in Fig.~\ref{fig:f2eps}.}\label{fig:alphamx}\end{center}
\end{figure}

For mass splittings $\delta \gtrsim 2 m_e$, the dark photon induced decay $\xs \to \x +2 e$ may deplete the remaining $\xs$ abundance to completely negligible levels~\cite{Finkbeiner:2009mi,Batell:2009vb}. However, for $\delta \ll m_e$, in the absence of an additional dipole-type interaction, the only kinematically allowed decays are $\xs \to \x + 3 \gamma$ and $\xs \to \x + 2 \nu$, with a corresponding lifetime that is cosmologically stable. In this case, the primordial $\xs$ fraction generically survives to late times, potentially giving rise to detectable signatures in cosmological and terrestrial observations. 

\subsection{Astrophysical Signatures}

Near the time of recombination, the primordial abundance of $\xs$ facilitates late time coannihilations to SM particles, depositing energy into the SM plasma and leading to small distortions in the CMB anisotropies. This process is suppressed by the small residual fraction $f_*$, but is compensated by the large number density of $\x$ for $m_\x \ll \text{GeV}$. The resulting energy injected into the SM plasma is strongly constrained by Planck observations, leading to $f_*\,  \sigma v \lesssim \text{pb} \times (m_\x / 60 \ \text{GeV})$ for electromagnetic final states~\cite{Aghanim:2018eyx}. The corresponding cross section for coannihilations to leptonic final states is 
\beq
\sigma v (\x \xs \to \ell \ell) \simeq \frac{16 \pi \, \alpha \, \alpha_D \, \epsilon^2 \, m_\x^2}{(4 m_\x^2-\mAp^2)^2}
~.
\eeq
The resulting Planck bound is shown in gray in Figs.~\ref{fig:f2eps}-\ref{fig:alphamx}. As shown explicitly in Fig.~\ref{fig:f2eps}, this constraint is strongest for $m_\x \sim \ \text{GeV}$, in which case $f_* \gtrsim \mathcal{O}(10^{-1})$. For much larger masses, $f_*$ saturates at $f_* \sim \mathcal{O}(1)$, while the DM number density falls as $\sim 1 / m_\x$, leading to a weakening of the bounds. 

Self-interactions in dark matter haloes and merging galaxy clusters constrain the rate for DM elastic scattering to be $\sigma (\x \x \to \x \x ) / m_\x \lesssim 10 \ \text{cm}^2 / \text{g}$~\cite{Tulin:2017ara}. Such limits restrict large values of $\alpha_D$ and are especially relevant at small DM masses, as shown in gray in Figs.~\ref{fig:f2eps} and \ref{fig:alphamx}. In scenarios with $f_* \ll 1$ and mass splittings greater than the typical DM kinetic energy, the dominant process at small masses arises from elastic scattering $\x \x \to \x \x$ that is radiatively induced by $\Ap$ exchange (see, e.g., Ref.~\cite{Berlin:2018jbm}). 

\subsection{Direct Detection}

The presence of a long-lived primordial $\xs$ component can also lead to signals in terrestrial direct detection experiments. In particular, if $\xs$ makes up a subcomponent of the galaxy's DM halo, down-scattering off of electrons $\xs e \to \x e$ leads to a mono-energetic recoil energy of  $E_R \simeq \mu \, \delta/ m_e$ provided that the mass splitting is greater than $\delta \gtrsim \mu \, v^2$, where $v$ is the $\xs$ velocity and the dark matter momentum is large compared to the typical electron momentum, see Sec.~\ref{sec:modelspace}. In the limit that $\delta \ll m_\x \ll \mAp$, the differential cross section for down-scattering is
\beq
\frac{d \sigma}{d E_R} \simeq \frac{8 \pi \, \alpha \, \alpha_D \, \epsilon^2 \, m_e}{\mAp^4 v^2}
~.
\eeq
At the level of our ``free'' electron approximation, the expected signal rate $R$ is then given by
\beq
\label{eq:PrimRate1}
R \equiv \frac{dN_{\text{sig}}}{dt \, dM_{\text{det}}} \simeq   \frac{N_A \, Z_{\text{free}}}{A \ \text{g}} \, \frac{f_* \, \rho_\x}{m_\x} \, \,  \frac{8 \pi \, \alpha \, \alpha_D \, \epsilon^2 \, m_e}{\mAp^4} \, \int_0^\infty \eta(E_R) \, dE_R \, \int_{v_\min}^\infty dv ~ \frac{f_{\text{halo}}(v)}{v}
~,
\eeq
where $\eta(E_R)$ is the detector efficiency as a function of recoil energy~\cite{Aprile:2020tmw}, $\rho_\x \simeq 0.3 \ \text{GeV} / \text{cm}^3$ is the local DM energy density, $v_\min \simeq |m_e E_R - \mu \delta| / (\mu \sqrt{2 m_e E_R})$ is the minimum kinematically allowed $\xs$ velocity, $M_{\text{det}}$ is the detector mass, $N_A$ is Avogadro's number, $Z_{\text{free}} = 26$  is the number of electrons in the $n = 4,5$ orbitals of xenon, and $A$ is the atomic mass. Approximating the halo velocity distribution $f(v)$ as Maxwellian with dispersion $v_0 \ll \sqrt{\delta / \mu}$, the recoil energy and velocity integrals reduce to
\beq
\int_0^\infty dE_R \, \int_{v_\min}^\infty dv ~ \frac{f_{\text{halo}}(v)}{v} \simeq \frac{(2 \mu)^{3/2} \delta^{1/2}}{m_e}
~,
\eeq
giving an overall event rate of
\beq
R \sim 10^6 \ \left( \text{tonne-year} \right)^{-1} \times f_* \, \Big( \frac{\delta}{\text{keV}} \Big)^{1/2} \Big( \frac{y}{10^{-10}} \Big) \Big( \frac{m_\x}{100 \ \text{MeV}} \Big)^{-5} 
~.
\eeq
Hence, even a very subdominant primordial fraction $f_* \ll 1$ can give rise to detectable rates. 

In Figs.~\ref{fig:epsmx} and \ref{fig:alphamx}, we highlight regions of parameter space in which an excited component of the DM energy density leads to electron down-scattering event rates at XENON1T, ranging from $(1-100) / (\text{tonne-year})$. As shown in Fig.~\ref{fig:f2eps}, larger $\alpha_D$ leads to a smaller primordial $\xs$ abundance, thus suppressing the down-scattering rate in terrestrial detectors. Also shown are regions excluded from recent missing energy/momentum searches at the low-energy accelerator experiments NA64 and BaBar~\cite{Lees:2017lec,NA64:2019imj}, as well as the projected sensitivities of a search for similar signals at LDMX and Belle II~\cite{Izaguirre:2014bca,Battaglieri:2017aum,Akesson:2018vlm,Berlin:2018bsc}.

Fig.~\ref{fig:epsmx} focuses on the $\epsilon-m_\x$ parameter space. In a standard cosmology, $\x$ freezes out via $\x \xs \leftrightarrow f f$ with an abundance consistent with the observed DM energy density along the black contours. Above or below these contours, $\x$ is a subdominant DM component or is overabundant assuming a standard cosmology. For concreteness, when calculating the signal even rate we take $\x$ to make up all of the DM throughout all of the parameter space shown. In Fig.~\ref{fig:alphamx}, $\epsilon$ is varied consistently in the $\alpha_D - m_\x$ plane such that $\x$ makes up all of the DM energy density. 
Regions in excess of $100 / \text{tonne-year}$ are constrained by a recently reported search for electron recoils in XENON1T~\cite{Aprile:2020tmw}. 

Assuming that thermal decoupling of $\x \xs \leftrightarrow f f$ sets the late-time $\x$ abundance, scenarios in which the ground state $\x$ makes up a subdominant component of the DM  lead to increasingly larger signal rates for $\xs$ down-scattering in terrestrial detectors. To see this, note that if $\x \xs \leftrightarrow f f$ decouples at temperatures much greater than $\delta$, then $f_\x \propto 1/(\alpha_D \, \epsilon^2)$, where $f_\x \equiv n_\x / n_{_{\text{DM}}} \leq 1$ is the DM fraction composed of $\x$. If the decoupling of $\xs \xs \leftrightarrow \x \x$ is responsible for setting the the $\xs$ abundance at much later times, then $f_* \propto 1/(\alpha_D^2 \, f_\x)$. The down-scattering signal rate at direct detection experiments is then controlled by the product $f_* f_\x \alpha_D \epsilon^2 \propto f_* \propto 1/(\alpha_D^2 f_\x)$. Hence, smaller $\x$ abundances imply larger signals in such cosmologies.

\section{Excited States from the Sun}
\label{sec:sun}

\subsection{Production}

When the excited state $\chi^*$ has a  lifetime  much shorter than the age of the universe due to the existence of, e.g., an electromagnetic dipole transition, the primordial abundance of $\chi^*$ can be severely depleted. In this case, detection of $\chi^*$ at direct detection experiments is only possible with a source of up-scattering. 

For a decay lifetime that is much larger than 1 AU divided by dark matter velocity, the Sun can act as a source of $\chi^*$. The Sun has a high internal temperature which we take to be $T_\odot = 1.1\, \kev$, and is capable of up-scattering the DM particles that come through it with $\sim \kev$ energies. Gravitational focusing due to the large gravitational field also enhances the flux of DM particles incident on the solar core. 

The idea to use ``reflected'' DM from the Sun was proposed in Ref.~\cite{An:2017ojc} in the context of elastic scattering. However, the reflected rates and energies are sufficiently low that  terrestrial experiments are typically more sensitive to the background primordial flux. This is not the case for inelastic WIMPs. For light WIMPs, even a small splitting $\delta \sim \, 100 \ \text{eV}$ can be kinematically inaccessible for up-scattering in a terrestrial experiment. Thus, {\em any} production in the Sun of an excited state which is suitably long-lived can produce a signal in a terrestrial experiment which would otherwise be absent.

To calculate the rate, we consider the problem as follows. In-falling DM particles in the core of the Sun have velocities $v \sim v_{\text{esc}} = 5 \times 10^{-3} \, c = 1500 \text{ km sec}^{-1}$, the escape velocity at the surface of the core. This is a high velocity compared to typical halo DM, $v_0 \sim 10^{-3} \, c$. The electrons in the sun are moving with an even higher velocity $v_e \sim \sqrt{2T/m_e}\sim 0.05 \, c = 1.5 \times 10^4\text{ km sec}^{-1}$. Since $v_{\text{esc}} \ll v_e$, we should think about the solar up-scattering with DM particles being essentially at rest, and being bombarded by thermal electrons from all around them. The quantity of interest is therefore the steady-state density of DM in the sun, and not the flux of DM on the sun. 

The ground state DM number density $n_{\chi,\odot}$ in the core of the sun is  enhanced by a factor $1+v_{\text{esc}}^2/v_{0}^2$ due to gravitational focusing. On the other hand, the higher velocity spreads the DM particles more thinly due to conservation of flux, suppressing the density by $v_{\text{esc}}/v_0$. Thus, we have $n_{\chi,\odot} \simeq n_0 \times v_{\text{esc}}/v_0$.\footnote{Precisely, the focusing is true for a $1/r$ potential. Inside the sun, this is no longer the case. However, approximately 50\% of the mass of the Sun is contained inside of $r<r_\odot/4$. Thus, we consider the 1/r potential to be reasonable down to these distances at the level of accuracy we have here.} The flux $\Phi$ of $\chi^*$ on Earth is then given by
\beq
	\Phi=  n_e\vev{\sigma_{\chi \to \chi^*} v_e}\times \frac{n_{\chi,\odot} V_\odot}{4 \pi (1 \text{ AU})^2} \,,
\eeq
where $\langle \sigma_{\chi \to \chi^*} v_e \rangle$ is the velocity-averaged cross-section of $\chi e^- \to \chi^* e^-$, and $V_\odot$ is the volume of the Sun's core. The derivative of the flux with respect to kinetic energy $K_{\chi^*}$ of the up-scattered $\chi^*$ is given by
\beq
    \frac{d\Phi}{dK_{\chi^*}} =  n_e\left<\frac{d\sigma_{\chi \to \chi^*}}{dK_{\chi^*}} v_e\right> \times \frac{n_{\chi,\odot} V_\odot}{4 \pi (1 \text{ AU})^2} \, .
    \label{eq:flux_per_energy}
\eeq
We take the solar parameters to be $V_\odot = 2.2 \times 10^{31} \text{ cm}^3$, and $n_e = 2 \times 10^{25} \text{ cm}^{-3}$ which is approximately the mean electron density in the solar core \cite{Bahcall:2000nu}. To get a sense of how large the flux is, we can compare it to the background flux of DM particles in the halo, $\Phi_0 = n_0 v_0$:
\begin{multline}
    \frac{\Phi}{\Phi_0} \simeq 5 \times 10^{-8} \left( \frac{n_e}{2 \times 10^{25} \text{ cm}^{-3}} \right)  \left( \frac{\langle \sigma_{\chi \to \chi^*} v_e \rangle}{10^{-30} \text{ cm}^3 \text{ s}^{-1}} \right)  \\
    \times\left( \frac{220 \text{ km/s}}{v_0} \right)\left( \frac{V_\odot}{2.2 \times 10^{31} \text{ cm}^3} \right) \left( \frac{v_{\text{esc}}/v_0}{7.0} \right) \,.
    \label{eq:solar_flux_ratio}
\end{multline}
For a benchmark value for $\langle \sigma_{\chi \to \chi^*} v_e \rangle = 10^{-30}\text{ cm}^{3} \text{ s}^{-1}$ that leads to an appreciable event rate in XENON1T, we find that the solar flux of $\chi^*$ is significantly smaller than the virial dark matter flux; however, the splitting $\delta$ greatly improves the detectability of $\chi^*$.

A simple expression for $\langle \sigma_{\chi \to \chi^*} v_e \rangle$ can be found in the nonrelativistic limit and with $\delta \ll m_e, m_\chi$, since the electron velocity distribution is Maxwellian. The differential scattering cross section in this limit is
\beq
    \frac{d \sigma_{\chi \to \chi^*}}{dK_{\chi^*}} = \frac{\overline{\sigma}_e m_\chi}{2 \mu_{\chi e}^2 v_e^2} \,,
\eeq
where $K_\chi$ is the recoil kinetic energy of $\chi$, and $\overline{\sigma}_e$ is defined in Eq.~\eqref{eq:DPxsec}. The velocity-averaged cross section is then
\beq
    \langle \sigma_{\chi \to \chi^*} v_e \rangle = \int_0^\infty dK_{\chi^*} \int_{v_\min}^\infty dv_e \, f_{\text{MB}}(v_e) \frac{d\sigma_{\chi \to \chi^*}}{dK_{\chi^*}} v_e \,,
    \label{eq:sigmave}
\eeq
where $v_{\text{min}}$ is the minimum velocity at fixed $K_\chi$ given by the kinematics of the up-scattering, 
\beq
    v_{\text{min}} = \frac{1}{\sqrt{2 m_\chi K_{\chi^*}}} \left(\frac{m_\chi K_{\chi^*}}{\mu_{\chi e}} + \delta \right) \,,
\eeq
and $f_{\text{MB}}(v_e)$ is the Maxwell-Boltzmann velocity distribution, 
\beq
    f_{\text{MB}}(v_e) = 4 \pi v_e^2 \left(\frac{m_e}{2 \pi T_\odot}\right)^{3/2} \exp \left(- \frac{m_e v_e^2}{2 T_\odot}\right) \,.
\eeq


The integrals in Eq.~(\ref{eq:sigmave}) can be evaluated analytically, giving
\begin{align}
	\langle \sigma_{\chi \to \chi^*}v_e \rangle &= \overline{\sigma}_e  \sqrt{\frac{2 m_e}{\pi T_\odot}} \frac{\delta}{\mu_{\chi e}}  \exp \left( -\frac{m_e \delta}{2 \mu_{\chi e} T_\odot} \right) K_1  \left( \frac{m_e \delta}{2 \mu_{\chi e} T_\odot} \right) \\
    &\simeq \overline{\sigma}_e \begin{cases}
         \sqrt{\frac{8 T_\odot}{\pi m_e}} \,, & \delta / \mu_{\chi e} \ll 2T_\odot / m_e \,,\\
         \sqrt{\frac{2\delta}{\mu_{\chi e}}} \exp \left(- \frac{m_e \delta}{\mu_{\chi e} T_\odot}\right) \,, & \delta / \mu_{\chi e} \gg 2T_\odot / m_e \, ,
    \end{cases}
    \label{eq:sigmave_analytic}
\end{align}
where we have expanded the Bessel function $K_1$ assuming a large argument for the final approximation. 

The factor of $\sqrt{2 \delta/\mu_{\chi e}}$ is a characteristic velocity of the up- and down-scattering process, with the exponential suppression coming from the fact that only electrons with $v_e \gtrsim \sqrt{2 \delta/\mu_{\chi e}}$ are capable of up-scattering $\chi$. Consequently, the cross section is suppressed exponentially compared to the elastic cross section if $\delta \gtrsim T_\odot$. 

However, we still have a somewhat surprising fact in that inelasticity benefits the signal tremendously. Ordinarily, the DM can only carry away $\sim \mu^2/m_\chi m_e\sim m_e/m_\chi$ fraction of the energy. However, because of the inelasticity, $\xs$s exit the sun with a substantial amount of energy to deposit in the detector. Thus, although the scattering rate is not increased compared to the elastic case, the {\em detectable} signal can be significantly enhanced.

\subsection{Direct Detection}

\begin{figure}\begin{center}
	\includegraphics[width=0.55\textwidth]{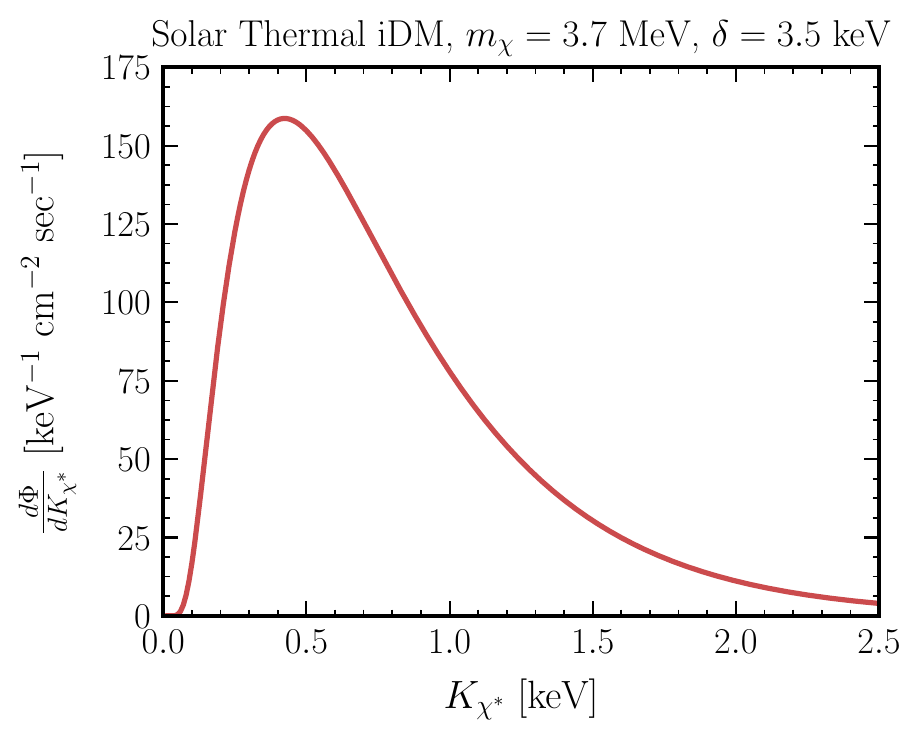}
	\caption{The flux $\Phi$ of $\chi^*$ particles up-scattered by electrons per unit kinetic energy $K_{\chi^*}$, assuming $m_\chi = 3.7 \, \mev$, $\delta = 3.5 \, \kev$ and a thermal annihilation cross section.}\label{fig:solarspectrum}\end{center}
\end{figure}

While primordial down-scatters yield a relatively narrow recoil electron spectrum in direct detection experiments at approximately the splitting $\delta$, the $\chi^*$ flux from the Sun has a broadened kinetic energy spectrum through scattering from thermal electrons; the rate per energy of $\chi^*$ particles produced for $m_\chi = 3.7 \, \mev$ thermal DM with a splitting of $\delta = 3.5 \, \kev$ is shown in Fig. \ref{fig:solarspectrum}.

{}

With the DM flux per energy $d \Phi/ dK_{\chi^*}$ in Eq.~\eqref{eq:flux_per_energy}, we can write the electron recoil spectrum per detector mass per time $dR/dE_R$ observed at a direct detection experiment as
\beq
    \frac{dR}{dE_R} = \frac{N_T}{M_{\text{det}}} \int dK_{\chi^*} \, \frac{d\Phi}{dK_{\chi^*}} \frac{d \sigma_{\chi^* \to \chi}}{dE_R} \,,
\eeq
where $E_R$ is the electron recoil energy, $\sigma_{\chi^* \to \chi}$ is the down-scattering cross section, $N_T$ is the number of targets in the detector, and $M_{\text{det}}$ is the detector mass. This expression can be evaluated numerically, but we can gain significant analytic understanding of $R$, the expected number of events per detector mass per time, at a direct detection experiment by assuming the nonrelativistic limit and $\delta \ll m_e,m_\chi$ once again. In this limit, the down-scattering cross section can be written in a particularly simple form:
\beq
    \frac{d \sigma_{\chi^* \to \chi}}{dE_R} \simeq \frac{\overline{\sigma}_e m_e}{2 \mu_{\chi e}^2 v_{\chi^*}^2} \,.
\eeq
The cross section of scattering for a given DM velocity $v_{\chi^*}$ can be obtained by integrating this expression up to the kinematic limit. In the elastic limit, this simply gives $\overline{\sigma}_e$; however, the existence of the splitting $\delta$ can extend this kinematic limit significantly, giving $\sigma_{\chi^* \to \chi} \simeq F \overline{\sigma}_e$, where following Eq.~\eqref{eq:kinenhance} we have
\beq
    F \equiv \sqrt{1 + \frac{2 \delta}{\mu_{\chi e} \langle v_{\chi^*}^2 \rangle}} \,,
\eeq
with $\langle v_{\chi^*}^2 \rangle$ the mean square velocity of $\chi^*$ from the sun; integrating over $d \Phi / dK_{\chi^*}$ shows that 
\begin{align}
    \langle v_{\chi^*}^2 \rangle &= \frac{\delta \mu_{\chi e}}{m_\chi} \frac{K_2 \left(\frac{\delta m_e}{2 \mu_{\chi e} T_\odot}\right)}{K_1 \left(\frac{\delta m_e}{2 \mu_{\chi e} T_\odot}\right)} \\
    &\simeq \begin{cases}
        \frac{8 \mu_{\chi e}^2 T_\odot}{m_\chi^2 m_e}\,, & \delta/\mu_{\chi e} \ll 2 T_\odot / m_e \,,  \\
        \frac{2 \delta \mu_{\chi e}}{m_\chi^2} \,, & \delta/\mu_{\chi e} \gg 2 T_\odot / m_e \,.
    \end{cases}
\end{align}
$F$ represents an enhancement with respect to the elastic scattering cross section, which is significant whenever the velocity scale $2\delta/\mu_{\chi e} \gg v_{\chi^*}^2$. With this result, we can write $R$ as
\beq
    R \simeq \frac{N_T}{M_{\text{det}}} \Phi F \overline{\sigma}_e \,.
\eeq
Combining the equation with the expression for the ratio of the solar flux to the DM halo flux in Eq.~\eqref{eq:solar_flux_ratio} and the analytic estimate for $\langle \sigma_{\chi \to \chi^*} v_e \rangle$ in Eq.~\eqref{eq:sigmave_analytic}, we obtain the following numerical estimate for $R$ for a xenon experiment in the solar inelastic DM model:
\begin{multline}
    R \simeq 26\, \left(\text{tonne-year}\right)^{-1}\left( \frac{n_e}{2 \times 10^{25} \text{cm}^{-3}} \right)  \left( \frac{\overline{\sigma}_e}{10^{-38} \text{ cm}^2 } \right)^2 \left( \frac{V_\odot}{2.2 \times 10^{31} \text{ cm}^3} \right) \left( \frac{v_{\text{esc}} / v_0}{7.0}\right) \\
     \times \left( \frac{\rho_0}{0.3 \text{ GeV cm}^{-3}} \right) \left(\frac{3.7\text{ MeV}}{m_\chi} \right) \left( \frac{F}{8.0} \right) \left( \frac{\sqrt{2\delta/\mu_{\chi e}} \exp[-m_e \delta/\mu_{\chi e} T_\odot]}{3 \times 10^{-3}} \right) \,,
     \label{eq:solar_rate_estimate}
\end{multline}
where $\rho_0$ is the local DM mass density. The values shown for comparison are either exactly the solar parameters adopted for our calculations, or are close to the actual values of these parameters when $m_\chi = 3.7 \, \mev$.

\begin{figure}\begin{center}
    \includegraphics[width=0.6\textwidth]{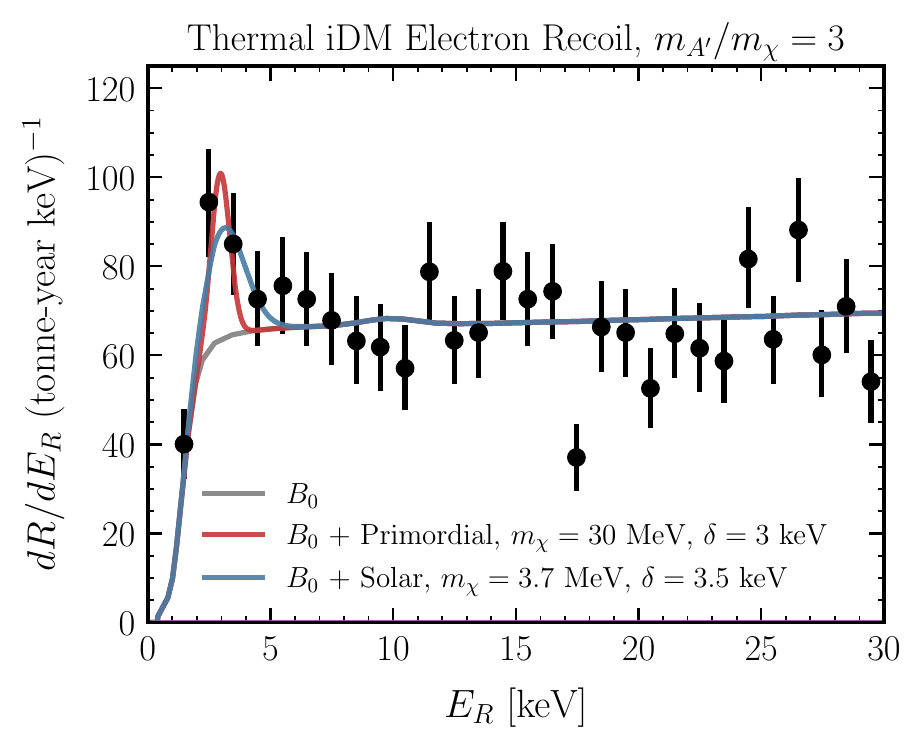}
    \caption{Detected electron recoil spectrum in the XENON1T experiment. We show the background model $B_0$ (gray) provided by Ref.~\cite{Aprile:2020tmw}, together with the $B_0$+signal for the thermal inelastic DM in the solar up-scattering with $m_\chi = 3.7\, \mev$, $\delta = 3.5 \, \kev$ (blue) and the primordial excited states scenarios with $m_\chi = 30 \, \mev$, $\delta = 3 \, \kev$ (red).}\label{fig:xesolarspec}\end{center}
\end{figure}

Armed with this analytic understanding, we are now ready to examine the numerical results. In Fig.~\ref{fig:xesolarspec}, we show an expected solar inelastic DM spectrum $dR/dE_R$ at XENON1T, together with the latest measurement of the event rate in the $(0-30) \ \text{keV}$ range and the experiment's background model~\cite{Aprile:2020tmw}. To obtain the signal spectrum, we convolve our result with the detector resolution~\cite{Aprile:2020yad} and multiply by the detector efficiency as a function of recoil energy~\cite{Aprile:2020tmw}. Here, we have chosen parameters that are consistent with a thermal inelastic DM model, with $m_\chi = 3.7 \, \mev$, $\mAp/m_\chi = 3$ and $\delta = 3.5 \, \kev$; these parameters lead to approximately 60 events per tonne-year at a xenon detector. Because the DM flux is generated by scattering with thermal electrons in the solar core, its kinetic energy spectrum significantly broader than the narrow dispersion expected from the primordial model.

\begin{figure*}[tbp]
    \centering
    \includegraphics[width=0.495\textwidth]{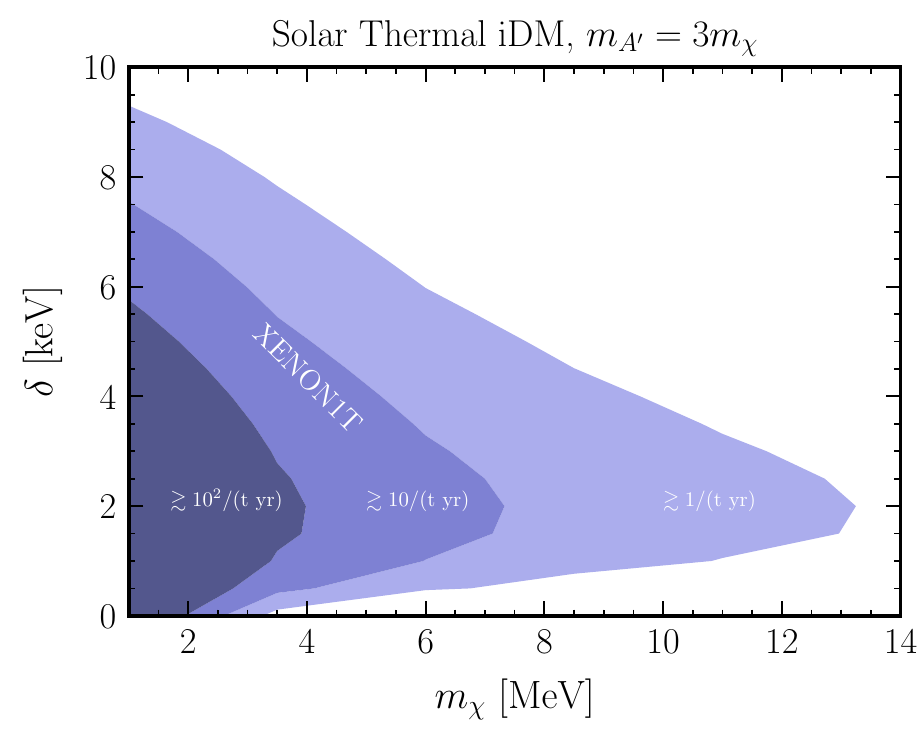}
    \includegraphics[width=0.495\textwidth]{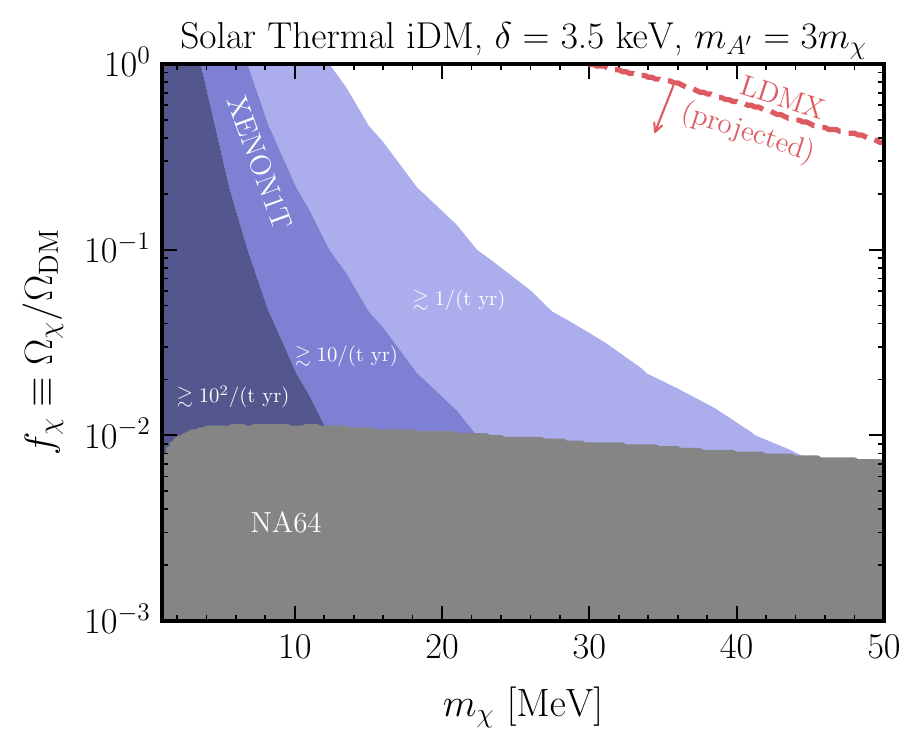}
    \caption{(Left) Expected event rate at XENON1T for the solar thermal inelastic DM model (blue), as a function of DM mass $m_\chi$ and the splitting $\delta$ (left) and as a function of $m_\chi$ and the thermal DM abundance by mass $f_\chi$ (right). Current limits from NA64~\cite{NA64:2019imj}(gray) as well as the future reach of LDMX~\cite{Akesson:2018vlm,Berlin:2018bsc} (red, dashed) are also shown. Note that the entire $m_\chi$--$\delta$ parameter space will be probed by LDMX. } 
    \label{fig:solar_thermal_delta_fchi}
\end{figure*}

Fig.~\ref{fig:solar_thermal_delta_fchi} (left) shows the expected rate $R$ at XENON1T as a function of the DM mass $m_\chi$ and the splitting $\delta$. For small splittings $\delta \ll T_\odot$, the enhancement in the down-scattering rate encoded in $F$ is close to 1, leading to a small rate. As the splitting increases to $\delta \sim \, \kev$, the enhancement becomes significant, and event rates of 100 per tonne-year can be expected for $m_\chi \lesssim 5$ MeV. Once $\delta \gg T_\odot$, however, few electrons in the solar core have sufficient energies to up-scatter $\chi$, leading to the exponential suppression shown in Eq.~\eqref{eq:solar_rate_estimate}. For a thermal model, since $\overline{\sigma}_e \propto \langle \sigma v \rangle_{\text{ann}} \mu_{\chi e}^2/m_\chi^2$ and for a sufficiently large splitting, $F \propto m_\chi$, we obtain $R \propto \delta e^{-m_e \delta/(\mu_{\chi e} T_\odot)} m_\chi^{-4} $, leading to a power law drop in $R$ as $m_\chi$ increases, and an exponential decrease as $\delta$ increases. There are currently no other experimental constraints in this range of parameters, but LDMX~\cite{Akesson:2018vlm,Berlin:2018bsc} will be sensitive to this entire parameter space.

Fig. 6 (right) shows a similar result but in the $m_\chi$--$f_\chi$ plane, where $f_\chi$ is the fractional mass abundance of $\chi$, which we assume to be thermally produced. Under this assumption,  $\rho_\chi \propto f_\chi$ and $\langle \sigma v \rangle \propto 1/f_\chi$, and so the overall rate at a direct detection experiment grows as $1/f_\chi$, making subdominant components easier to detect. A similar argument as before gives $R \propto f_\chi^{-1} m_\chi^{-4}$, so that lines of constant event rate on the $m_\chi$--$f_\chi$ plane follows $f_\chi \propto m_\chi^{-4}$. XENON1T can probe thermal iDM through solar scattering of all abundances below 10 MeV for $\delta \sim 3 \, \kev$. Other constraints on this plane include the NA64 experiment~\cite{NA64:2019imj}, which has ruled out all sub-100 \mev\, thermal dark matter with $f_\chi \lesssim 0.01$, and the future LDMX experiment~\cite{Akesson:2018vlm,Berlin:2018bsc} which probes a similar parameter space to XENON1T. 

\begin{figure*}[tbp]
    \centering
    \includegraphics[width=0.495\textwidth]{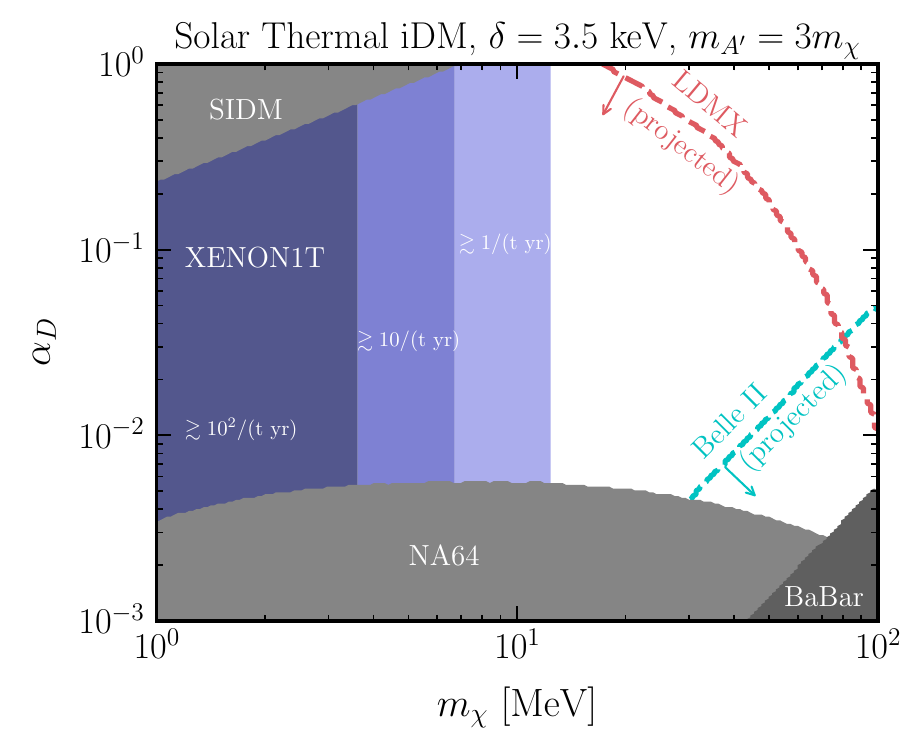}    
    \includegraphics[width=0.495\textwidth]{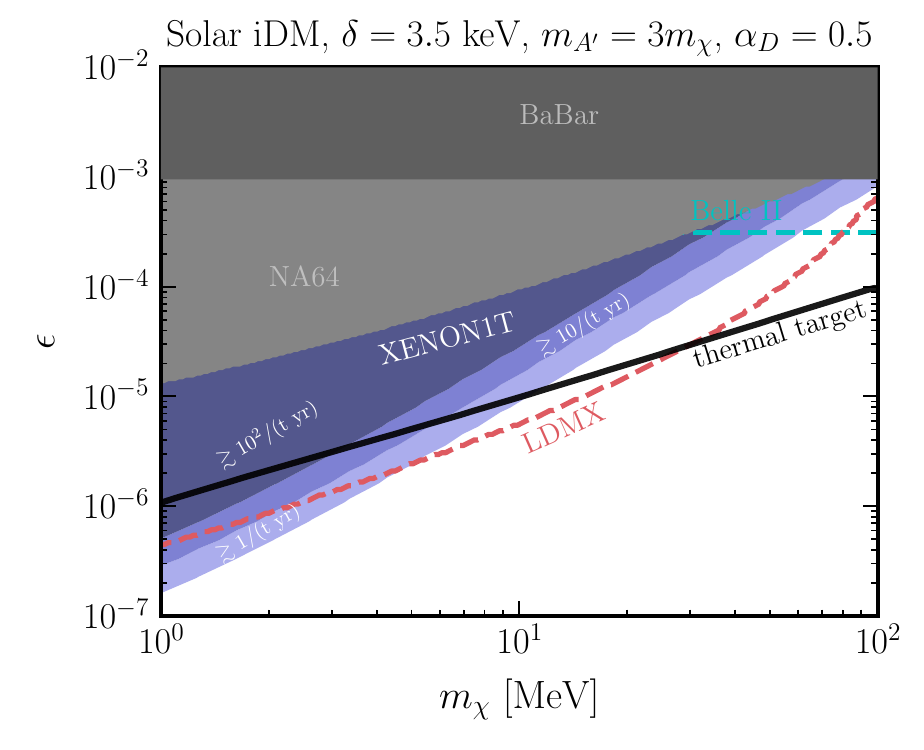}
    \caption{Expected event rate at XENON1T for the solar inelastic DM model (blue), as a function of $m_\chi$ and $\alpha_D$ \textit{assuming} thermal production (left), and as a function of $m_\chi$ and $\epsilon$ \textit{without} assuming thermal production, fixing $\alpha_D = 0.5$ (right). Current constraints from NA64~\cite{NA64:2019imj}, BaBar~\cite{Lees:2017lec} and self-interaction of DM~\cite{Tulin:2017ara} are shown in gray, with the future reaches of Belle II~\cite{Izaguirre:2015zva,Battaglieri:2017aum} (green, dashed) and LDMX~\cite{Akesson:2018vlm,Berlin:2018bsc} (red, dashed) displayed in both plots. } 
    \label{fig:solar_eps_alpha}
\end{figure*}

In Fig.~\ref{fig:solar_eps_alpha}  (left), we consider the $m_\chi$--$\alpha_D$ plane for a thermally produced dark matter with $f_\chi = 1$. In this plane, the rate does not depend on $\alpha_D$ since  $\langle \sigma v \rangle_{\text{ann}}$ is held approximately constant by varying $\epsilon$. Once again, we see the relation $R \propto m^{-4}$. At higher DM masses $m_\chi \gtrsim 10 \, \mev$, current bounds are limited to $\alpha_D \lesssim 10^{-2}$, but Belle II~\cite{Izaguirre:2015zva,Battaglieri:2017aum} and LDMX can potentially probe $10^{-2} \lesssim \alpha_D \lesssim 1$ up to $m_\chi = 1 \, \gev$.

Finally, in Fig.~\ref{fig:solar_eps_alpha}  (right), we lift the assumption of thermal DM, and  assume that $\chi$ makes up all of the dark matter through an unspecified non-thermal production mechanism. After fixing the coupling $\alpha_D = 0.5$,  we show the region of the $\epsilon-m_\x$ parameter space where we expect 1--100 events per tonne-year in XENON1T, as well as existing and future beam dump constraints. Without the thermal dark matter assumption, the choice of $m_{A'} / m_\chi = 3$ means that $\overline{\sigma}_e \propto m_\chi^{-8}$, so that overall the event rate at XENON1T scales as $R \propto \epsilon^4 m_\chi^{-8}$. For this  choice of mass splitting, we can see that xenon direct detection experiments have the potential to probe the thermal target line up to $m_\chi \sim 20 \, \mev$, with a reach comparable to that of the future LDMX. 

\subsection{Astrophysical Signatures}

Constraints from indirect detection and the CMB power spectrum do not apply to the parameter space of inelastic DM up-scattered in the Sun; the excited state is assumed to decay and is completely depleted over cosmological timescales, making annihilation of DM into Standard Model particles negligible. Self-interaction limits of 10 cm$^2$ g$^{-1}$~\cite{Tulin:2017ara} do place constraints at large values of $\alpha_D$ and small values of $m_\chi$, as shown in  Fig.~\ref{fig:solar_eps_alpha} (\textit{left}).

Other potential indirect signals include photon emission after excitation in other high-temperature environments. Dark matter up-scattering  in hot gas followed by a decay through a dipole has been proposed as an explanation \cite{Finkbeiner:2014sja,DEramo:2016gqz} for potential astrophysical excesses in X-ray spectra at 3.5 keV \cite{Bulbul:2014sua,Boyarsky:2014jta}. 
The X-ray flux from the Perseus cluster from DM excitation is estimated to be~\cite{DEramo:2016gqz}
\beq
\Phi \simeq 10^{-5}\ {\rm sec^{-1}\  cm^{-2}}\times\left(\frac{\mev}{m_\chi}\right)\left(\frac{\rho_\x}{\rho_{\mathrm{DM}}}\right)\left(\frac{\vev{\sigma_{\x\,e} v}}{10^{-24}\ \rm cm^3 \ sec^{-1}}\right), 
\eeq
to be compared to the observed flux of around $10^{-5}{\rm \ sec^{-1} \ cm^{-2}}$. 

In our model, the excitation is mediated by the dark photon interaction; a similar calculation to that of the Sun, but using a hot gas temperature of $T=6.8\  \kev$  \cite{1992MNRAS.254} yields $\vev{\sigma v} \simeq 4 \times 10^{-29}\  {\rm cm^3 \ sec^{-1}}$ for $m_\chi = 4\mev$ and $\delta=3.5 \, \kev$, several orders too low for the putative signal. Moreover, the expected rate is the same for  a thermal abundance regardless of the dark matter fraction, as increasing the cross section  decreases the relic abundance. Nonetheless, the possibility is intriguing and we leave a detailed study for future work.

\section{Excited States from the Earth}
\label{sec:earth}

Finally, we consider excited states with the shortest lifetimes, which can be populated locally by dark-photon-mediated up-scattering in the Earth.  Subsequent electromagnetic decays can yield a detectable signal in parts of parameter space where the scattering process itself is currently unobservable. These classes of `luminous' models have been considered in the context of dipole up-scattering in the Earth \cite{Feldstein:2010su,Eby:2019mgs}, in material near the target \cite{Pospelov:2013nea}, and in the detector itself \cite{Chang:2010en,Lin:2010sb}. 

Unlike in the hot environments in the Sun and the early universe, the relative velocities on Earth are too low for DM-electron scattering to populate splittings on the scale of a keV, so we focus here on nuclear scatterings.  Furthermore, the DM must have sufficient mass to kinematically up-scatter at typical DM velocities: for $\delta \sim \kev$ one must have $m_\chi \gtrsim \gev$ to scatter without kinematical suppression. 

In this mass range, the thermal relic makes up the full DM abundance for  $y\sim 10^{-8}$, with a DM-proton scattering cross section of $\sigma_{p} \simeq 2 \times 10^{-37}{\rm cm^2}\left( \frac{\rm GeV}{m_\chi}\right)^2$
in the elastic limit. While the elastic nuclear recoil cross section of this magnitude is excluded by CRESST \cite{Abdelhameed:2019hmk},  a splitting $\delta$ produces a kinematic suppression $f_{\text{inel}}$ in the scattering rate, avoiding existing limits. The suppression  has steep sensitivity to the masses and halo parameters, ranging from $f_{\text{inel}}\sim10^{-2}-10^{-5}$ for $\kev\lesssim\delta \lesssim 2\  \kev$ at $\mx\simeq\gev$, and $f_{\text{inel}}\sim10^{-1}-10^{-7}$ for $\kev\lesssim\delta \lesssim 2.5 \ \kev$ at $\mx\simeq1.2\ \gev$. Particles with $m_\chi = 3 \, \gev$ and $f_{\text{inel}}\sigma_p \sim 10^{-42}\rm cm^2$, for instance, would evade current constraints. 

Each volume unit in the Earth acts as a source of up-scattered states and generates a local flux at a detector.  Considering the Earth as composed of crust, mantle and core, and the scatterings dominated by silicon and iron densities as in  Ref.~\cite{Eby:2019mgs}, we find the resulting flux of up-scattered excited states relative to the DM flux is given by
\beq
\Phi_{*}/\Phi_{\rm DM} \sim \frac{f_{\rm inel} \,\sigma_p}{5 \times10^{-34} \cm^2},
\eeq
At splittings close to the kinematic threshold, the flux is further enhanced because the up-scattered states have lower average velocity than that of the DM. 

Because the cross sections are proportional to the reduced mass of the system, the electron down-scattering cross section is suppressed by a factor $\sim m_e^2/m_\chi^2 \sim 10^{-6}$. However, for lifetimes long compared to the time to traverse the Earth $R_E/v_0$, decays can produce significant event rates in direct detection experiments. As previously discussed, a lifetime of $100 \sec$ allows the entire Earth to act as a source and requires a dipole suppressed $\Lambda_d \gtrsim 100 \tev$.

The rate per unit mass in XENON1T is approximately,
\beq
R\sim\frac{60}{\textrm{ tonne-year}}\left(\frac{f_{\rm inel} }{10^{-5}}\right)\left(\frac{\sigma_p}{10^{-37}\rm cm^2}\right)\left(\frac{\gev}{m_\chi}\right)\left(\frac{\rm 100\, sec}{\tau}\right)
\eeq
It is noteworthy that a thermal relic can naturally give a detectable rate in the $\gev$ mass range. Moreover, as before, for a subdominant thermal component, the increased cross section and decreased abundance will cancel, keeping the rate constant. We thus conclude that a thermal relic is capable of yielding a photon signal in direct detection experiments, with a precise rate prediction requiring further study.

The excited states propagate outside the Earth and can decay, producing a diffuse X-ray background peaked at the splitting energy. As the states are generated in the Earth, the flux of decaying states falls off as $1/r^2$ and the X-ray signal is dominated by the excited DM particles closest to the Earth. The rate is directly proportional to the volumetric rate ${dN}/{dtdV}$ in a DM direct detection experiment, yielding a flux of
\beq
\Phi_{\rm diffuse} \sim 0.1 \frac{\rm photons}{\rm \ sr \ cm^2 \, sec}\left(\frac{{dN}/{dtdV}}{10^4\rm \ m^3 \ year}\right)
\eeq
The limits are $O(0.1) {\rm \ sr^{-1} \ cm^{-2} \ sec^{-1}}$ in this energy range \cite{moretti2012spectrum}, making current terrestrial detectors more sensitive than X-ray satellites, as long as the decay length is large compared to the Earth radius; detections at future direct detection experiments could be correlated with diffuse X-ray signals.

Furthermore, DM-DM scattering can give rise to a population of excited states which then decay, again giving rise to a potential excesses in X-ray spectra  \cite{Finkbeiner:2014sja}, with flux comparable to the flux of Perseus for large enough cross sections,
\beq
\Phi_{\rm Perseus}  \simeq 10^{-5}{\rm \ sec^{-1}\  cm^{-2}}\times\left(\frac{\gev}{m_\chi}\right)^2\left(\frac{\vev{\sigma_{\x\,\x} v}}{10^{-21}\rm \ cm^3 \ sec^{-1}}\right).
\eeq
For GeV-mass DM considered here, we find $\vev{\sigma v} \sim 10^{-21} {\rm \ cm^3 \ sec^{-1}}$, potentially of the right order to source the tentative signal.

\section{Discussion}
\label{sec:discussion}
Models of light dark matter are simple and viable and result in a new class of experimental signatures. For light fermions coupled to a dark photon, CMB constraints naturally point to a pseudo-Dirac class of models.  These models come with an excited state that is often swept aside in the discussion of the DM phenomenology. In this paper, we have found that far from being a side note, these excited states can offer powerful signatures of this class of DM models. The fact that the scenarios we have considered may explain the putative excess at XENON1T adds to the excitement.

We have investigated three separate scenarios: primordial excitations,  excitations in the Sun, and excitations in the Earth. Each of them probes different regions of parameter space, and provides different implications for future experiments.

For primordial excitations, we have showed that the abundance of the excited state is typically  exponentially suppressed, with excited fractions as small as $f_* \sim 10^{-9}$  for light ($m_\chi \sim \mev$) thermal dark matter. We find this is true for splittings over a wide range of $\sim\kev - 100\  \kev$. Heavier ($m_\chi \sim \gev$) particles see a less pronounced, $\mathcal{O}(1)$ suppression of the excited state abundance. Note that this is quite unlike previous scenarios with heavy DM particles, where often $\chi$ and $\chis$ are present in roughly equal abundances. Remarkably, subdominant thermal DM components, i.e., when $\rho_\x<\rho_{\text{DM}}$, are even {\em more} constrained by experimental searches as they typically have a higher excited state abundance. 

The suppression of the excited state abundance naturally changes the signal rate. Nonetheless, existing and upcoming liquid xenon experiments exclude parts of  light dark matter parameter space. A direct detection signal would manifest as narrow line, except at DM masses close to the electron mass where the bound electron momentum broadens the recoil energy spectrum. In particular, we find that for thermal relic DM and $\alpha_D\gtrsim 10^{-2}$---bounded below by beam-dump and CMB constraints---these scenarios predict in excess of 100 events/tonne/year at a xenon experiment. For larger but still perturbative values of $\alpha_D$, rates remain excess of 1 event/tonne/year, making future Xenon experiments capable of testing much of the remaining parameter space. A large fraction of the parameter space will also be probed by LDMX and Belle II.  

For cases where the primordial states are unstable on cosmological timescales, the early universe abundance cannot contribute to a direct detection signal. However, local up-scatterings offer promise over a narrow, yet interesting, parameter space.

The Sun efficiently up-scatters light dark matter when the energy splitting is of order the temperature of the solar core. This allows dark matter to carry energy from the Sun and deposit it into terrestrial experiments. For a thermal relic making up all of the dark matter, one can expect detectable rates up to splittings as large as $10 \ \kev$ and masses up to $13 \ \mev$. For subdominant components, the scattering rate in the Sun remains constant as the scattering cross section increases, thus the flux of excited states at Earth does not decrease for subdominant DM components (until the Sun becomes opaque to them). However, the scattering cross section goes up and we again find direct detection experiments are more sensitive to subdominant components of dark matter. We find interesting signal rates up to masses of $\sim 50\  \mev$. This entire parameter space should be tested by LDMX.

For $\sim \gev$ masses, DM can up-scatter via the dark photon in the Earth and decay via photon emission. There is a narrow window at the $\gev$ scale where one can up-scatter at a detectable rate without conflicting with existing nuclear recoil experiments. Such a rate may also be detectable by future X-ray satellites.

Given the recent claim of an excess of electron events at XENON1T, it is exciting to consider these three scenarios as possible sources; all three make concrete predictions for future experiments. Future datasets from liquid Xenon experiments may be able to differentiate the energy spectra of the line shape predicted by the primordial abundance and the Earth up-scattering scenario versus the thermally broadened signal from solar up-scattering.

All three scenarios require an excited state near 3.5 keV to explain the XENON1T data. Intriguingly, there have been claims of excess X-ray emission from a variety of astrophysical sources in this range. For up-scatters in the Earth from a dark photon, the natural size of the cross section is adequate to explain the Perseus excess. For the lighter, solar up-scattered model, the X-ray flux is too small but remains an intriguing possibility. Conversely, some models that can explain the 3.5 keV line may be constrained by our analysis.

In summary, we have considered the electromagnetic signals arising from excited states, a  generic feature in viable models of light fermionic dark matter. We find the presence of these excited states leads to signals which already constrain the parameter space and provides exciting possibilities for discovery in current and future direct detection experiments. The same parameter space will be testable in future laboratory experiments such as LDMX and Belle II. 

\acknowledgments{}

We thank Rouven Essig and Siddharth Mishra-Sharma for helpful discussions and providing code and insights into DM-electron scattering; Natalia Toro and Mariana Carrillo Gonz\'{a}lez for detailed discussions and comparison to their work;  and Ken Van Tilburg for helpful discussion and comments on the manuscript.  AB and MB are supported in part by the James Arthur Fellowship. HL is supported by the DOE under contract DESC0007968 and the NSF under award PHY-1915409. NW is supported by NSF under award PHY-1915409 and the Simons Foundation. This research made use of the \texttt{IPython}~\cite{PER-GRA:2007}, \texttt{Jupyter}~\cite{Kluyver2016JupyterN}, \texttt{matplotlib}~\cite{Hunter:2007}, \texttt{NumPy}~\cite{numpy:2011}, \texttt{seaborn}~\cite{seaborn}, \texttt{SciPy}~\cite{2020SciPy-NMeth}, and \texttt{tqdm}~\cite{da2019tqdm}  software packages.

\bibliography{esignals}
\bibliographystyle{JHEP}
\end{document}